\documentclass[useAMS,usenatbib]{mn2e}
\usepackage{times}
\usepackage{graphics,epsf}
\usepackage{amsmath}                
\usepackage{amsfonts}               
\usepackage{amssymb}                
\usepackage{epsfig}                 
\usepackage{rotating}
\usepackage{color}
\usepackage{graphics,epsf}
\usepackage{amsmath}                
\usepackage{amsfonts}               
\usepackage{amssymb}                
\usepackage{epsfig}                 
\usepackage{subfigure}
\usepackage{float}
\usepackage{url}
\usepackage{times}
\usepackage{graphics,epsf}
\usepackage{amsmath}                
\usepackage{amsfonts}               
\usepackage{amssymb}                
\usepackage{epsfig}                 
\usepackage{rotating}
\usepackage{color}
\def \cm{~\rm{cm}}
\def \s{~\rm{s}}
\def \km{~\rm{km}}

\def \g{~\rm{g}}

\def \erg{~\rm{erg}}
\def \yrs{~\rm{yrs}}
\def \yr{~\rm{yr}}
\def \pc{~\rm{pc}}

\def \apj{ApJ}

\def \mnras{MNRAS}
\def \apjl{ApJ Lett.}
\def \apjs{ApJ Suppl. Ser.}

\begin{document}

\title{Type Ia Supernova Remnants: Shaping by Iron Bullets}

\author[Danny Tsebrenko and Noam Soker]
{Danny Tsebrenko and Noam Soker \\
Department of Physics, Technion -- Israel Institute of Technology, Haifa
32000, Israel; \\ ddtt@tx.technion.ac.il; soker@physics.technion.ac.il}

\maketitle

\begin{abstract}
Using 2D numerical hydrodynamical simulations of type Ia supernova remnants (SNR Ia)
we show that iron clumps few times denser than the rest of the SN ejecta might form protrusions in an otherwise spherical SNR.
Such protrusions exist in some SNR Ia, e.g., SNR~1885 and Tycho.
Iron clumps are expected to form in the deflagration to detonation explosion model.
In SNR Ia where there are two opposite protrusions, termed `ears’, such as Kepler's SNR and SNR~G1.9+0.3, our scenario implies that the dense clumps,
or iron bullets, were formed along an axis. Such a preferred axis can result from a rotating white dwarf progenitor.
If our claim holds, this offers an important clue to the SN Ia explosion scenario.

\end{abstract}


\maketitle

\section{Introduction}
\label{sec:intro}

Type Ia Supernovae (SNe Ia) are thermonuclear detonations of carbon-oxygen white dwarfs (WDs;{ }\citealt{Hoyle1960}).
Six SN Ia scenarios are currently considered, with no consensus on even the leading scenario for SN Ia.
(a) {\it The core-degenerate (CD) scenario} (e.g., \citealt{Livio2003, Kashi2011, Soker2011, Ilkov2012, Ilkov2013, Soker2013}).
Here the WD merges with a hot core of a massive asymptotic giant branch (AGB) star. Explosion might occur shortly or a long time after merger.
(b) {\it The double degenerate (DD) scenario} (e.g., \citealt{Webbink1984, Iben1984}).
According to this scenario a merger of two WDs takes place,
but there is no specification of the later evolution, e.g., how long after merger explosion occurs.
(e.g., \citealt{Pakmor2013, vanKerkwijk2010, Levanonetal2015, Tanikawaetal2015})
(c) {\it The 'double-detonation' (DDet) mechanism} (e.g., \citealt{Woosley1994, Livne1995, Shenetal2013}),
in which a sub-Chandrasekhar mass WD accumulates a layer of helium-rich material on its surface.
The helium layer detonates and leads to a second detonation near the center of the CO WD.
A variation of this scenario is the triple detonation scenario \citep{Papishetal2015}.
(d) {\it The single degenerate (SD) scenario} (e.g., \citealt{Whelan1973, Nomoto1982, Han2004}).
In this scenario the WD accretes mass from a non-degenerate stellar companion and explodes more or less when it reaches the Chandrasekhar mass limit.
(e) {\it The WD-WD collision (WWC) scenario} (e.g. \citealt{Kushniretal2013}).
In this scenario, two WDs collide and immediately ignite.
(f) The singly-evolved star (SES) scenario \citep{Chiosietal2015}.
In this scenario it is claimed that pycnonuclear reaction might be able to drive powerful detonations in single CO white dwarfs.

A comparison between the first five various
scenarios is given in \cite{TsebrenkoSoker2015a} and \cite{Soker2015}.
\cite{Papishetal2015} list two problems with the SES scenario.
When confronting the scenarios with observations, as well as when trying to account for a specific observational property, it is imperative to consider all scenarios.

In the present study we focus on the observational feature of two opposite 'ears' in some SN Ia remnants (SNRs Ia).
An `ear' is defined as a protrusion from an otherwise round SNR surface.

In a previous paper \citep{TsebrenkoSoker2013} we showed that opposite ears observed in some SNRs Ia
might be attributed to two opposite jets working either a long time before the explosion
or very shortly (within hours) before the SN explosion.
In the first possibility the ears are formed in the circumstellar matter (CSM) that was ejected a long time before explosion,
much as ears formed in planetary nebulae (PNe).
The general structure of a SN Ia inside a PN was mentioned by \cite{DickelJones1985}, although they did not refer to ears or to jets.
The formation of ears in this case of SN Ia inside PNe (SNIP; \citealt{TsebrenkoSoker2015a})
was simulated before for the Kepler SNR  \citep{TsebrenkoSoker2013}
and SNR~G1.9+0.3  \citep{TsebrenkoSoker2015b}.
This type of interaction can occur in the SD or CD scenarios.
But if the CSM is massive, $\ga 0.1 M_\odot$, then only the CD scenario can work.
In \cite{TsebrenkoSoker2013} we also simulated jets launched by an accretion disk
around the exploding WD very shortly before the explosion.
Such disks can be formed from the merger of two WDs in the DD scenario
(e.g., \citealt{Yoonetal2007, LorenAguilar2009, Jietal2013, Zhu2013}), or in the CD scenario.
Another possibility to form a small ear in the SNR is by the instabilities arising in the ejecta \citep{Warren2013}.

In the current paper we continue our study of ears in SNRs,
but study their formation by dense clumps, or 'bullets', along an axis inside the ejecta from the SN.
{{{ We would like to isolate the effects of iron bullets formed in the explosion from  possible  effects of dense CSM.
We therefore differ from our previous works on ears formation resulting from ejecta-CSM interaction \citep{TsebrenkoSoker2013, TsebrenkoSoker2015b},
and study an SN explosion inside a low-density ISM/CSM.}}}

We are motivated by the observations of iron clumps along two almost perpendicular axes in the SNR~1885 \citep{Fesenetal2014},
and by the presence of heavy metals inside the ears in Tycho \citep{Chiotellisetal2013}.
We note though that in Tycho's SNR the ears are not two opposite ears.
Such nickel/iron clumps might form during the deflagration phase of the deflagration to detonation (DTD)
scenario \citep{Seitenzahletal2013}.
\cite{Seitenzahletal2013} performed 3D simulations of the delayed-detonation explosion model,
and found the stable neutron-rich iron-group isotopes at intermediate velocities in a shell surrounding a nickel-rich core.
We are motivated by this finding as well, but we note that the recent analysis by \cite{Mazzalietal2015}
of the first year spectral evolution of the SN~2011fe does not support the claim for
neutron-rich iron group isotopes moving outside the nickel core.
{{{ 
SN ejecta clumps in core collapse SNRs have been simulated by \cite{Micelietal2013}, explaining observed protrusion features in the Vela SNR.
The role of ejecta clumping in forming protrusions in SNRs Ia has been studied also by \cite{Orlandoetal2012} for SNR~1006.
}}}
We perform 2D hydrodynamical simulations of an SN Ia explosion {{{ already}}} having an iron clump or a 'bullet' inside the SN ejecta,
and follow the evolution of the SNR.

The 2D numerical code and initial setting and conditions are described in Section \ref{sec:numerical}.
In Section \ref{sec:results} we show that the 'ears' in the otherwise round shapes of SNRs Ia
can be explained by the presence of dense iron clumps in the outer ejecta.
Our discussion and a short summary are given in Section \ref{sec:summary}.

\section{NUMERICAL SETUP}
\label{sec:numerical}
The simulations are performed using the high-resolution multidimensional hydrodynamics code {\sc{flash 4.2.2}} \citep{Fryxell2000}.
We employ a 2D adaptive mesh refinement grid having axisymmetric cylindrical $(x,y)$ geometry,
with the $y$-axis serving as the rotational symmetry axis.
The length of each axis is $\Delta = 20 \pc$.
The SN ejecta density $\rho_{\rm{ejecta}}$ is modelled by an exponential density profile
\citep{Dwarkadas1998},
\begin{equation}\label{eq1}
\rho_{\rm{ejecta}} = A \,{\rm{exp}}{} (-v/v_{\rm{ejecta}})t^{-3},
\end{equation}
where $v_{\rm{ejecta}}$ is a constant which depends on the mass and kinetic energy of the ejecta,
\begin{equation}
v_{\rm{ejecta}} = 2.44 \times 10^8 E_{51}^{1/2} \left(\frac{M_{\rm{SN}}}{M_{\rm{Ch}}}\right)^{-1/2} \cm \s^{-1},
\end{equation}
$M_{\rm Ch} = 1.4 M_\odot$,
$E_{51}$ is the explosion energy in units of $10^{51} \erg$, and $A$ is a parameter given by
\begin{equation}
A = 7.67 \times 10^6 \left(\frac{M_{\rm{SN}}}{M_{\rm Ch}}\right)^{5/2} E_{51}^{-3/2}  \g \s^{3} \cm^{-3} .
\end{equation}
We take the parameters as in \cite{TsebrenkoSoker2013, TsebrenkoSoker2015b}:
exploding mass of $M_{\rm Ch} = 1.4 M_\odot$, explosion energy of $10^{51} \erg$, and maximum ejecta velocity of $v_{\rm SNm}=20,000 \km \s^{-1}$.

\begin{figure}
\begin{center}
\includegraphics*[scale=0.33,clip=true,trim=0 0 0 0]{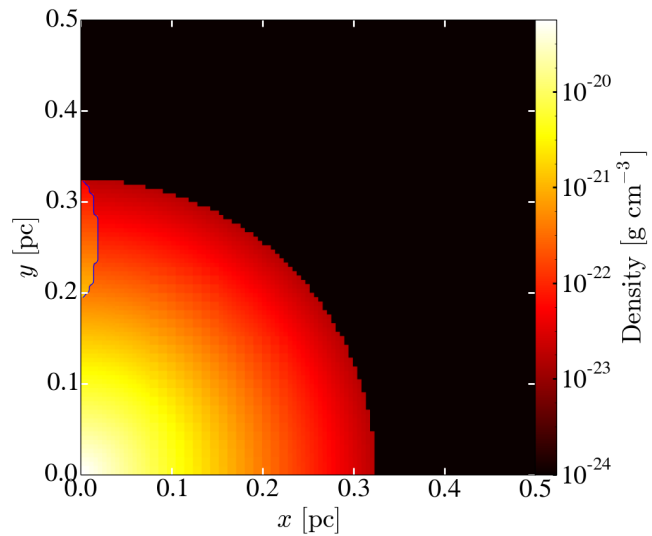}
\caption{The initial geometry of the supernova ejecta and the iron clump.
Color represents the density according to the color-bar on the right.
The initial iron clump density at each radius  is taken to be 3 times the density of the ejecta at the same radius.
Blue thin line marks the surface of the clump.
The total mass in one clump is $\simeq 0.01 M_\odot$. The mass of the ejecta inside a similar volume would be a third of that mass. }
\label{fig:initial}
\end{center}
\end{figure}

Our simulations start around ${{{ 16 \yrs}}}$ after the SN explosion.
By this time, the SN ejecta has reached a distance of $0.32 \pc$ from the explosion location.
The mass of the iron clump is taken to be $\simeq 0.01-0.02 M_\odot$,
and its density $\rho_{\rm{blob}}$ is taken to be 1.5-3 times the density of the adjacent SN ejecta, $\rho_{\rm{ejecta}}$ (equation \ref{eq1}).
The length of the clump is taken to be $0.13 \pc$ and its width (twice the distance from the symmetry axis) is taken to be $D = 0.02-0.03 \pc$.
{{{  We note that \cite{Orlandoetal2012} populate the SN ejecta with many much
smaller clumps having size of $0.01-0.02 \pc$ and then let the ejecta evolve, starting the simulation $10 \yrs$ after the SN explosion.
We start with a smooth SN ejecta having a single clump on each side along the symmetry axis (we simulate only one side).
}}}
We take the  initial ISM density outside the front of the ejecta to be $ \rho_{\rm ISM}= 10^{-24} \g \cm^{-3} $.
The initial density setup is presented in Figure \ref{fig:initial}, and the initial location of the iron blob is marked by a thin blue line.
The initial velocity of the material in the clump is identical to the velocity of the adjacent SN ejecta at the same radial distance,
which expands radially with velocities up to $v_{\rm SNm}$.

\section{RESULTS}
\label{sec:results}
\begin{figure*}
\begin{center}
\subfigure{\label{subfigure:Density1}\includegraphics*[scale=0.3,clip=true,trim=0
0 0 0]{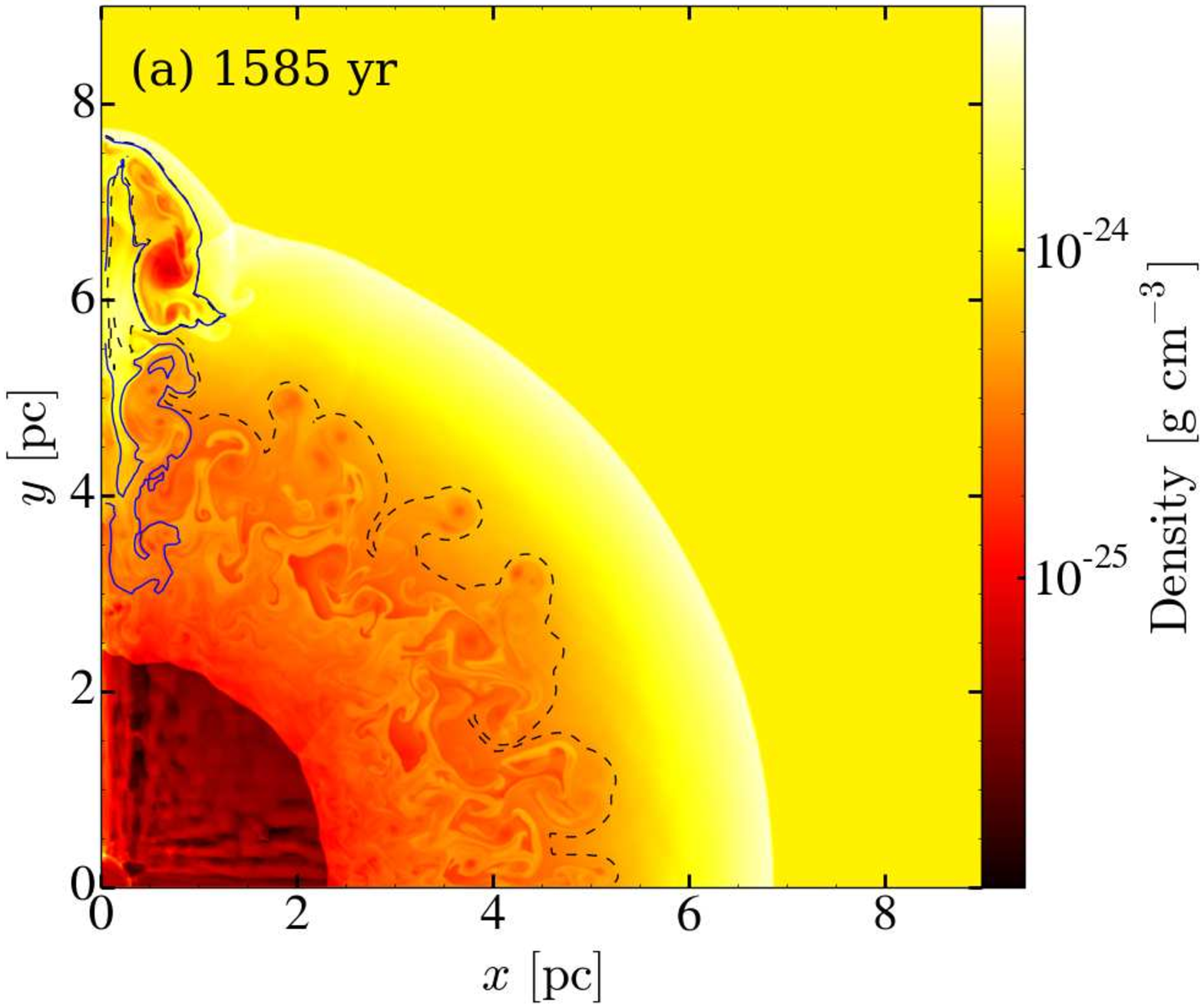}}
\subfigure{\label{subfigure:Temperature1}\includegraphics*[scale=0.3,clip=true,trim=0
0 0 0]{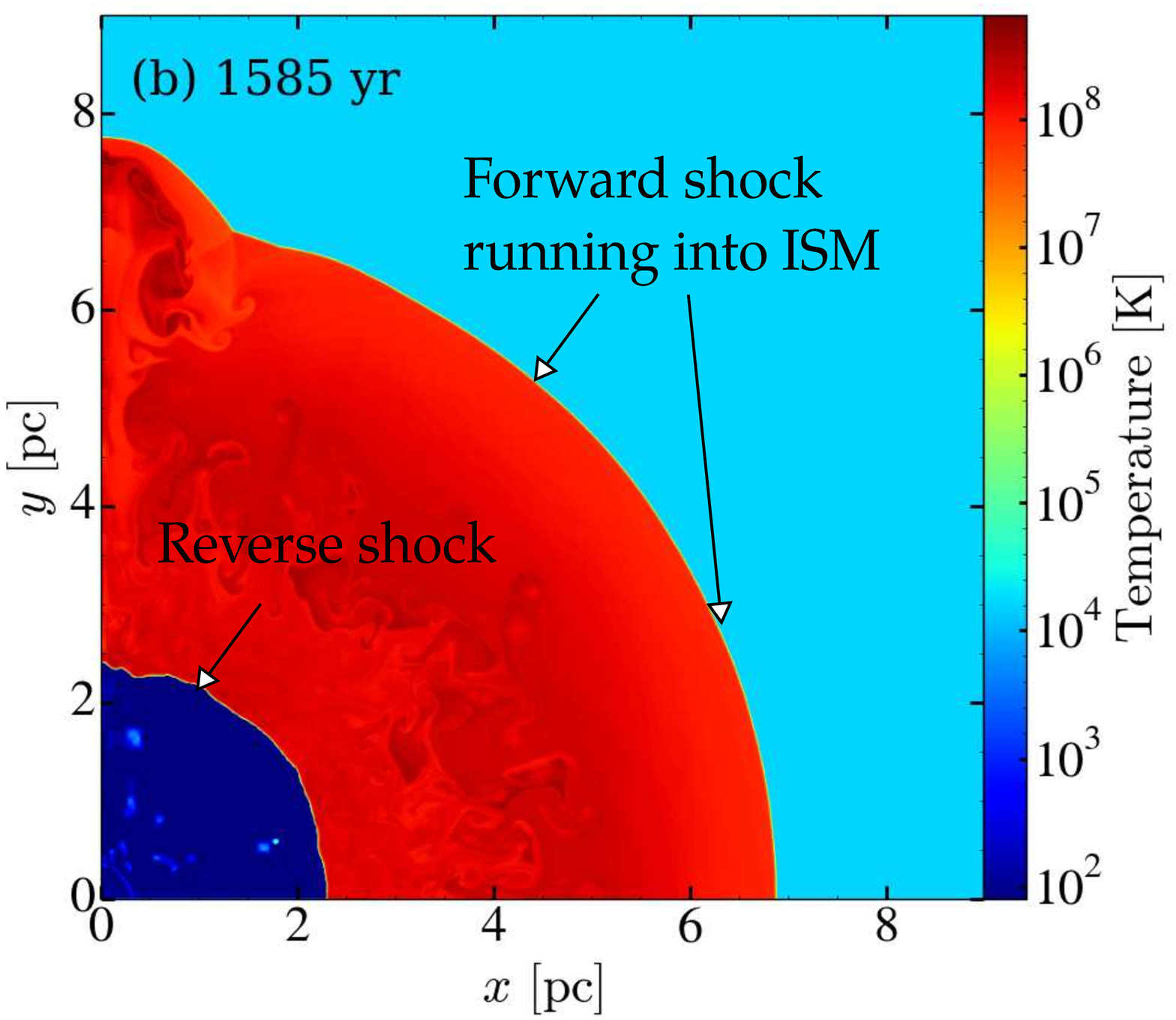}}
\subfigure{\label{subfigure:Pressure1}\includegraphics*[scale=0.3,clip=true,trim=0
0 0 0]{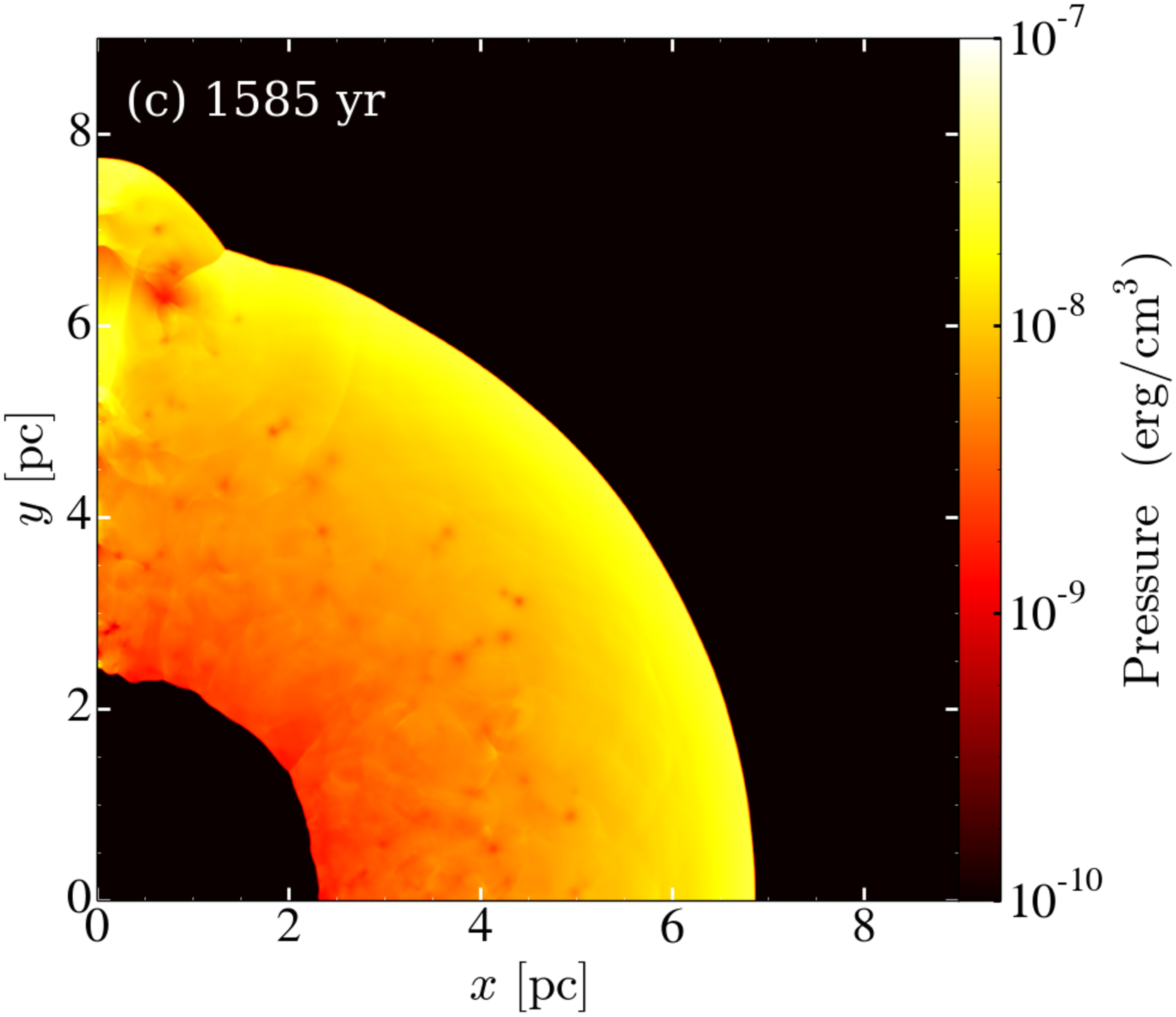}}
\subfigure{\label{subfigure:Magv1}\includegraphics*[scale=0.3,clip=true,trim=0
0 0 0]{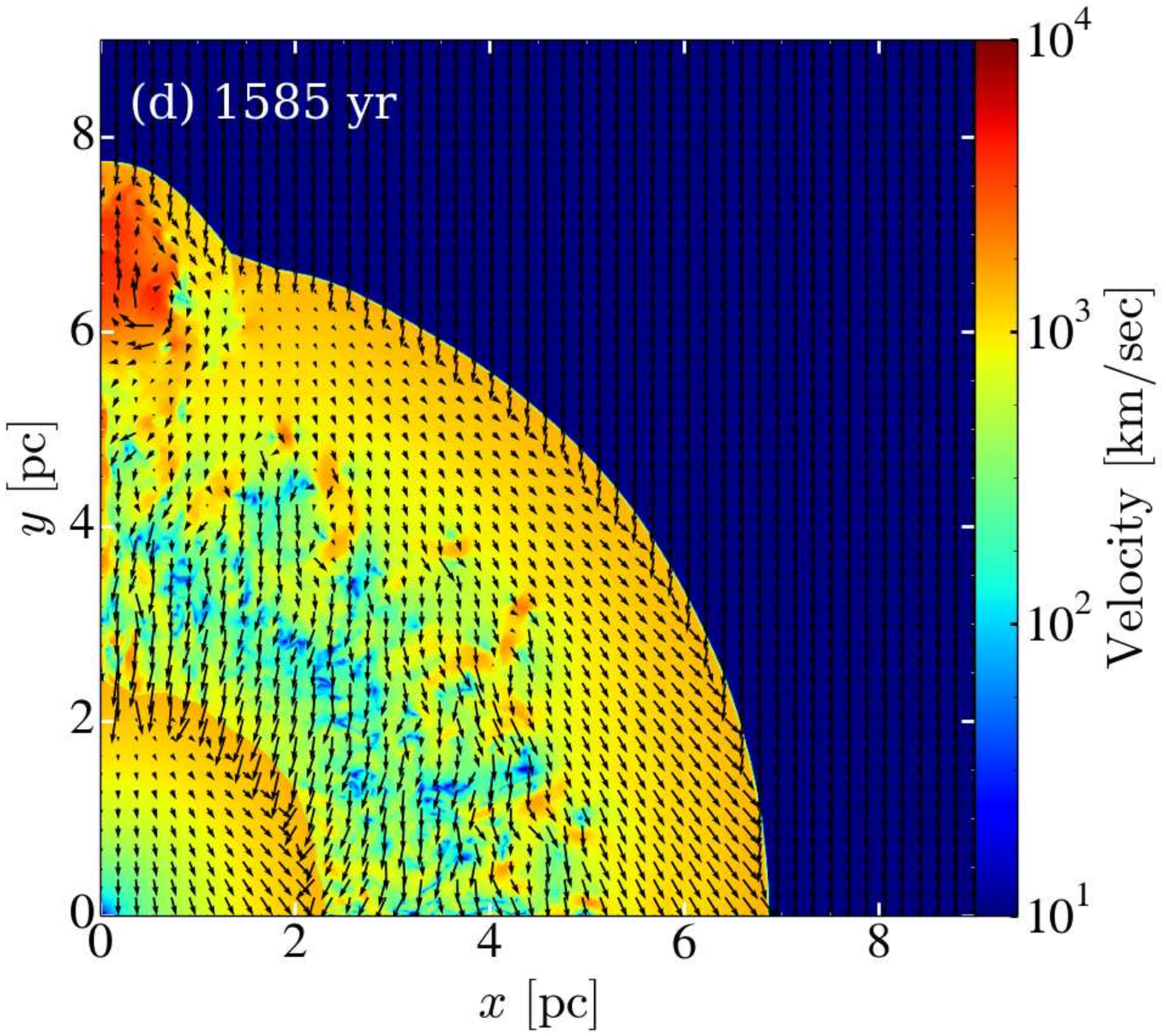}} \caption{ The flow properties at $t=1585 \yrs$
for our fiducial case with a blob width of $D = 0.02 \pc$ and blob
density of $\rho_{\rm blob} (r) = 3 \rho_{\rm ejecta}(r)$, whose
initial setting is shown in Fig. \ref{fig:initial} . (a) Density
map. The blue contour line shows the location of material
initially located within the iron clump. The dashed black contour
line marks the contact discontinuity, i.e., the boundary between
the shocked ejecta and the chocked ISM. (b) Temperature; (c)
Pressure; (d) Velocity magnitude according to the color scheme.
The flow direction and magnitude as given by the black arrows are
calculated after subtracting $200 \km \s^{-1}$ from the x velocity
component and $1,600 \km \s^{-1}$ from the y velocity component.
This was done to emphasize the vortex present around $(0.2, 6)
\pc$. }
 \label{fig:results}
\end{center}
\end{figure*}

As the SN ejecta interacts with the ISM, part of the kinetic energy of the ejecta is deposited in the interstellar medium (ISM)
and the ISM is accelerated outward, as the SN ejecta is decelerated. The denser iron blob continues to advance into the ISM,
and forms an ear in the shape of the resulting SNR.
Two shock waves are formed, one running outward into the ISM and one inward into the incoming ejecta.
A contact discontinuity develops between the shocked ejecta and the shocked ISM.
These, as well as various physical parameters,
for our fiducial run are presented in Figure \ref{fig:results}.
Panel \ref{subfigure:Density1} shows the density structure, in which the iron clump material is outlined by a thin blue line.
The time shown is years after simulation starts.
We present results at $t=1585 \yrs$ when the forward shock has reached a radius of $r_{\rm shock} =6.9 \pc$ (beside the iron clump that is further out).
The total swept-up ISM mass is $M_{\rm ISM}(t = 1585)\simeq 20 M_\odot$. The ejecta has been substantially decelerated. The ear feature is clearly apparent.

The relevant flow properties are as follows.
$(i)$ Instabilities.
As the ISM decelerates the ejecta a pressure gradient with
decreasing pressure toward the center is developed within the
ejecta. The density on the other hand increases from the ISM
toward the center. This situation is prone to Rayleigh-Taylor (RT)
instability. Indeed, well developed RT fingers and blobs are
clearly seen in the density and temperature maps. However, these
do not change the over all shape of the SNR and its boundary with
the undisturbed (pre-shock) ISM.
The instabilities are similar to the results obtained by \cite{Warren2013}.

 $(ii)$ As expected, the denser
iron blob deceleration is lower, and it penetrates the ISM.
An Ear structure has emerged, {{{ having mostly material originally in the Fe bullet.
This feature resembles morphologically the FeL emitting Ear structure in Tycho's SNR \citep{Chiotellisetal2013}.
A quantitative comparison between this feature in our simulation and the one in the
X-ray image of Tycho’s SNR will require further numerical simulation and deeper observational analysis.
These are beyond the scope of this exploratory study.  
}}}
{{{  Our result is similar to the protrusion obtained by the simulation of the Vela~SNR performed by \cite{Micelietal2013}.
The ear feature in our simulation is much larger than the many slight perturbations to SNR sphericity obtained by \cite{Orlandoetal2012}.
}}}
 $(iii)$ A vortex has developed relative to
the rest frame of the moving iron clump. Such a vortex can
increase the survivability of the clumps as it is being
decelerated by the ISM, making the Ear a long lasting structure.
To emphasize the vortex we have subtracted a velocity of $(v_x,
v_y) = (1600 , 200) \km \s^{-1}$ from all points. The color coding
in the velocity panel gives the absolute magnitude of the velocity
before this subtraction.

\begin{figure*}
\begin{center}
\subfigure{\label{subfigure:Density0}\includegraphics*[scale=0.3,clip=true,trim=0 0 0 0]{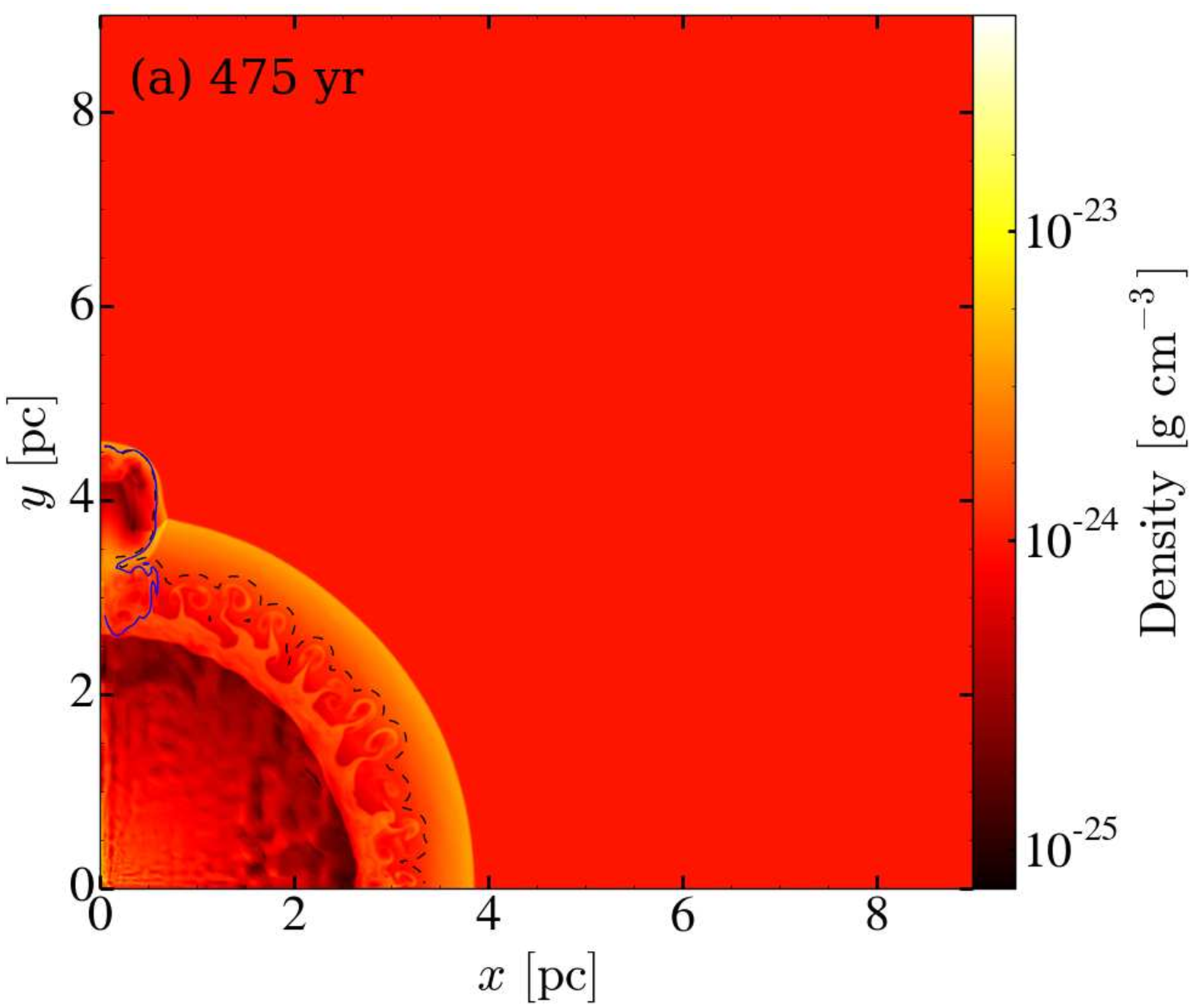}}
\subfigure{\label{subfigure:Temperature0}\includegraphics*[scale=0.3,clip=true,trim=0 0 0 0]{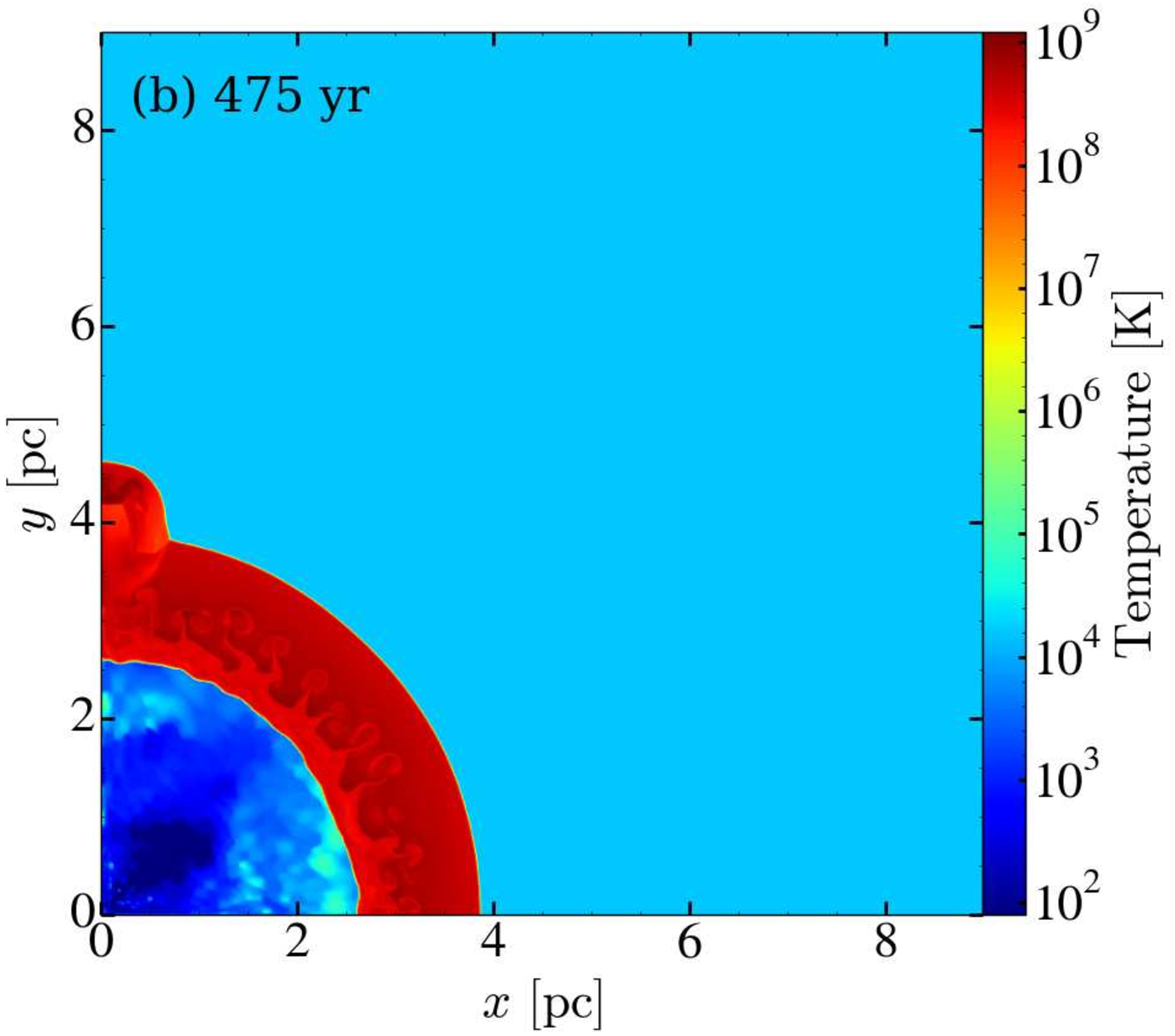}}
\subfigure{\label{subfigure:Pressure0}\includegraphics*[scale=0.3,clip=true,trim=0 0 0 0]{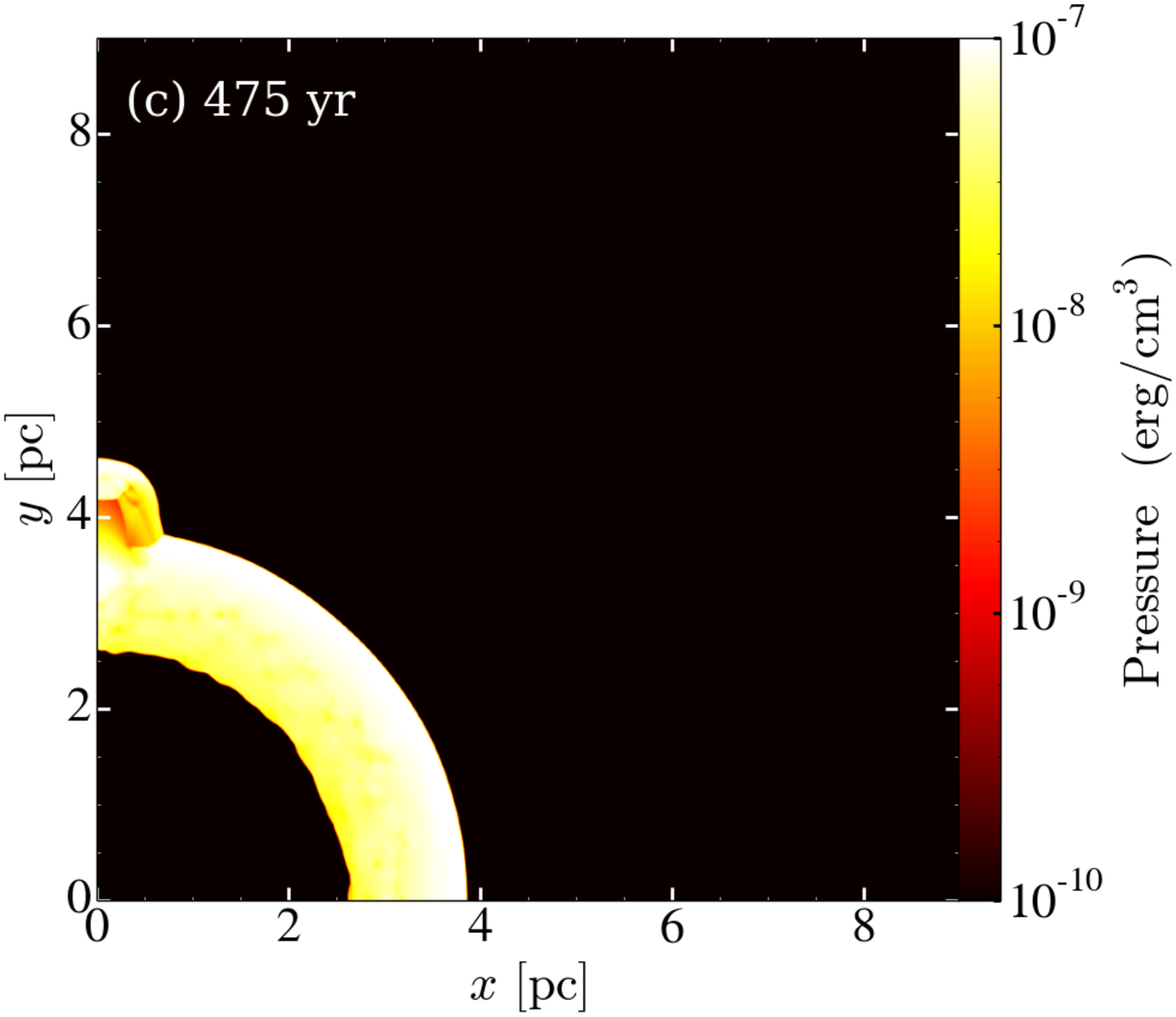}}
\subfigure{\label{subfigure:Magv0}\includegraphics*[scale=0.3,clip=true,trim=0 0 0 0]{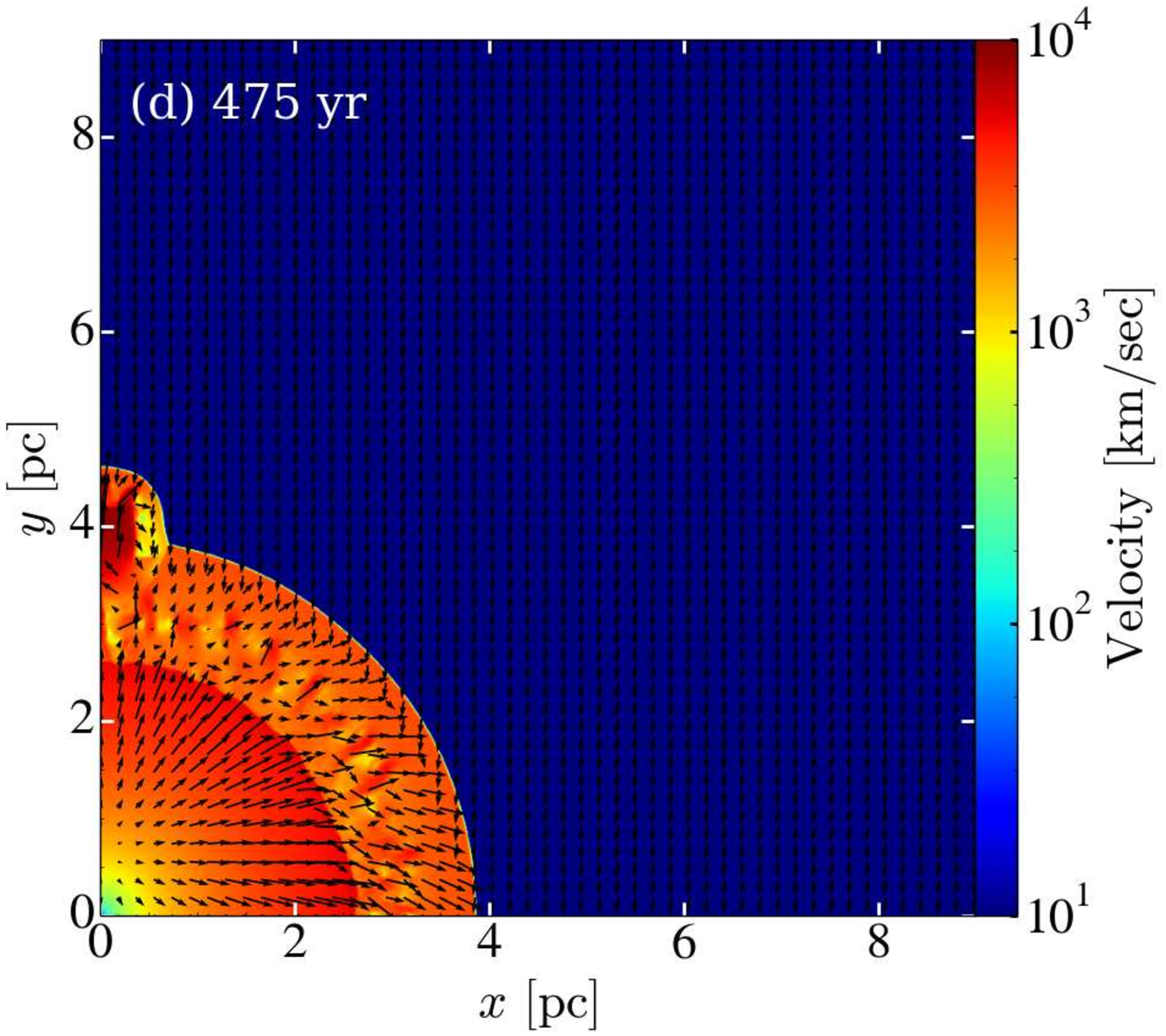}}
\caption{Same as Figure \ref{fig:results}, but at an earlier time (t=$475 \yr$). }
\label{fig:results0}
\end{center}
\end{figure*}

Figure \ref{fig:results0} shows our fiducial run at an earlier
evolution time, t=$475 \yr$. By this time the mass of the swept
ISM is comparable with the SN ejecta mass. The ear feature is
distinctly visible at this early evolution time as well.

\begin{figure*}
\begin{center}
\subfigure{\label{subfigure:Density2}\includegraphics*[scale=0.3,clip=true,trim=0 0 0 0]{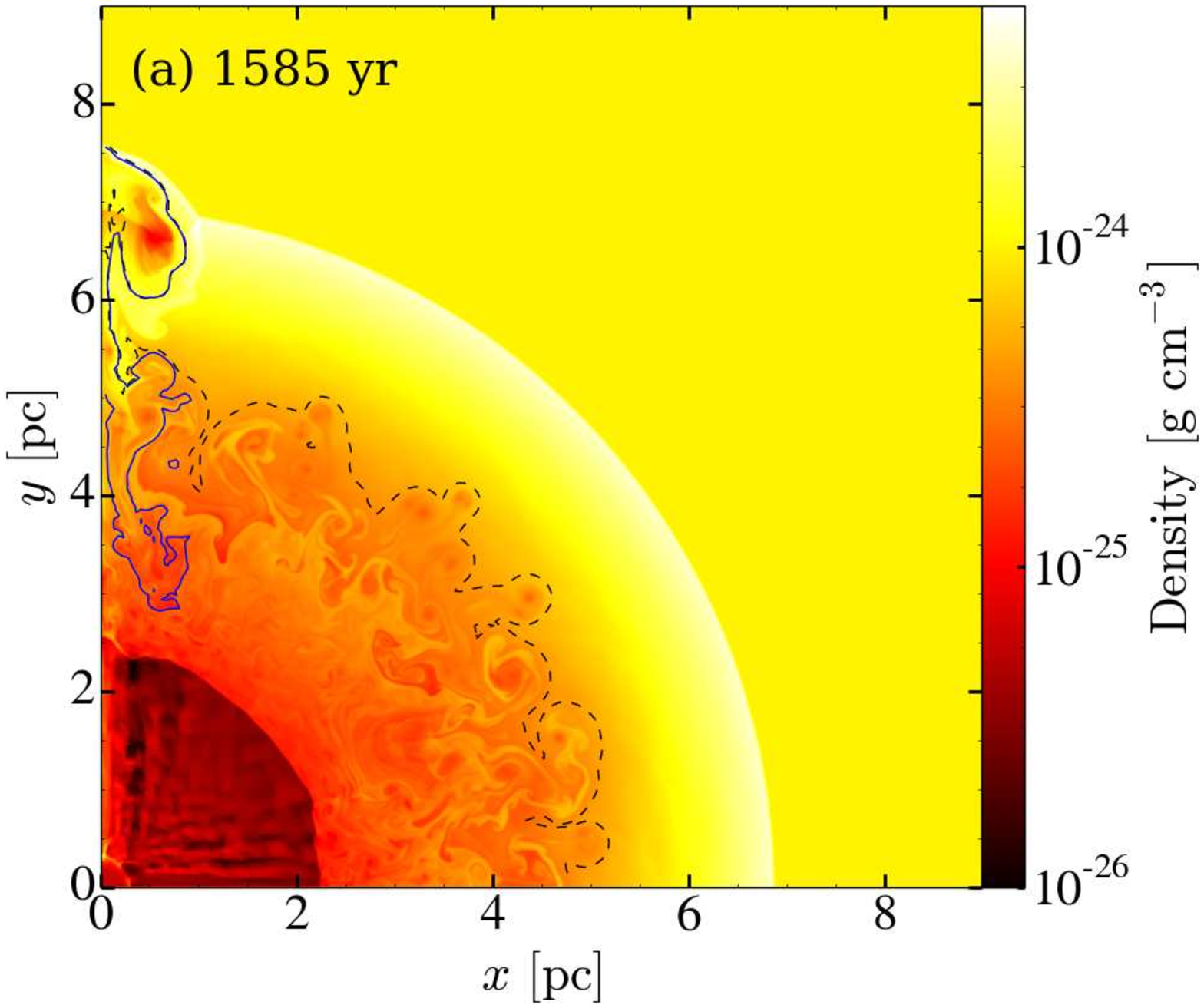}}
\subfigure{\label{subfigure:Temperature2}\includegraphics*[scale=0.3,clip=true,trim=0 0 0 0]{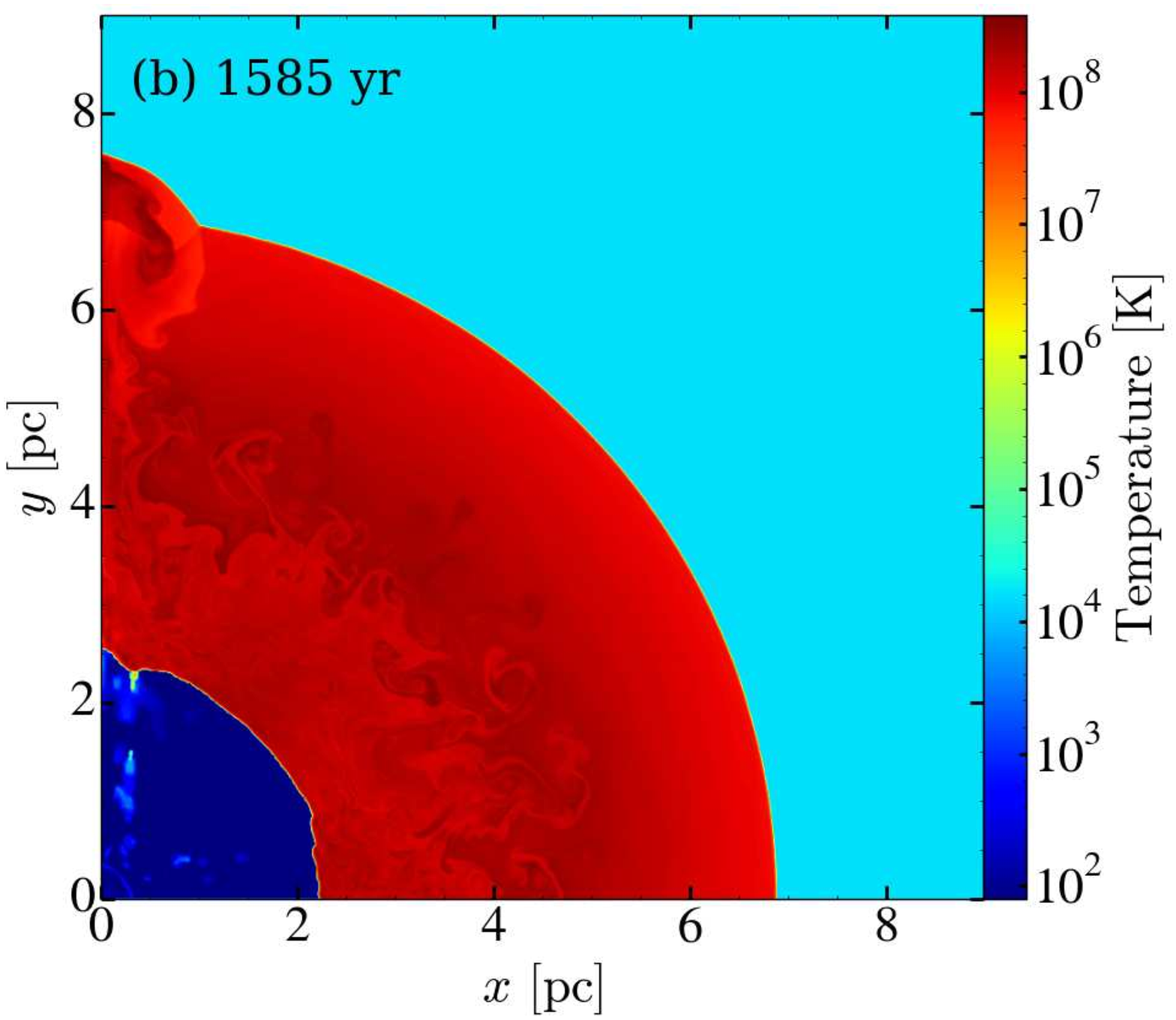}}
\subfigure{\label{subfigure:Pressure2}\includegraphics*[scale=0.3,clip=true,trim=0 0 0 0]{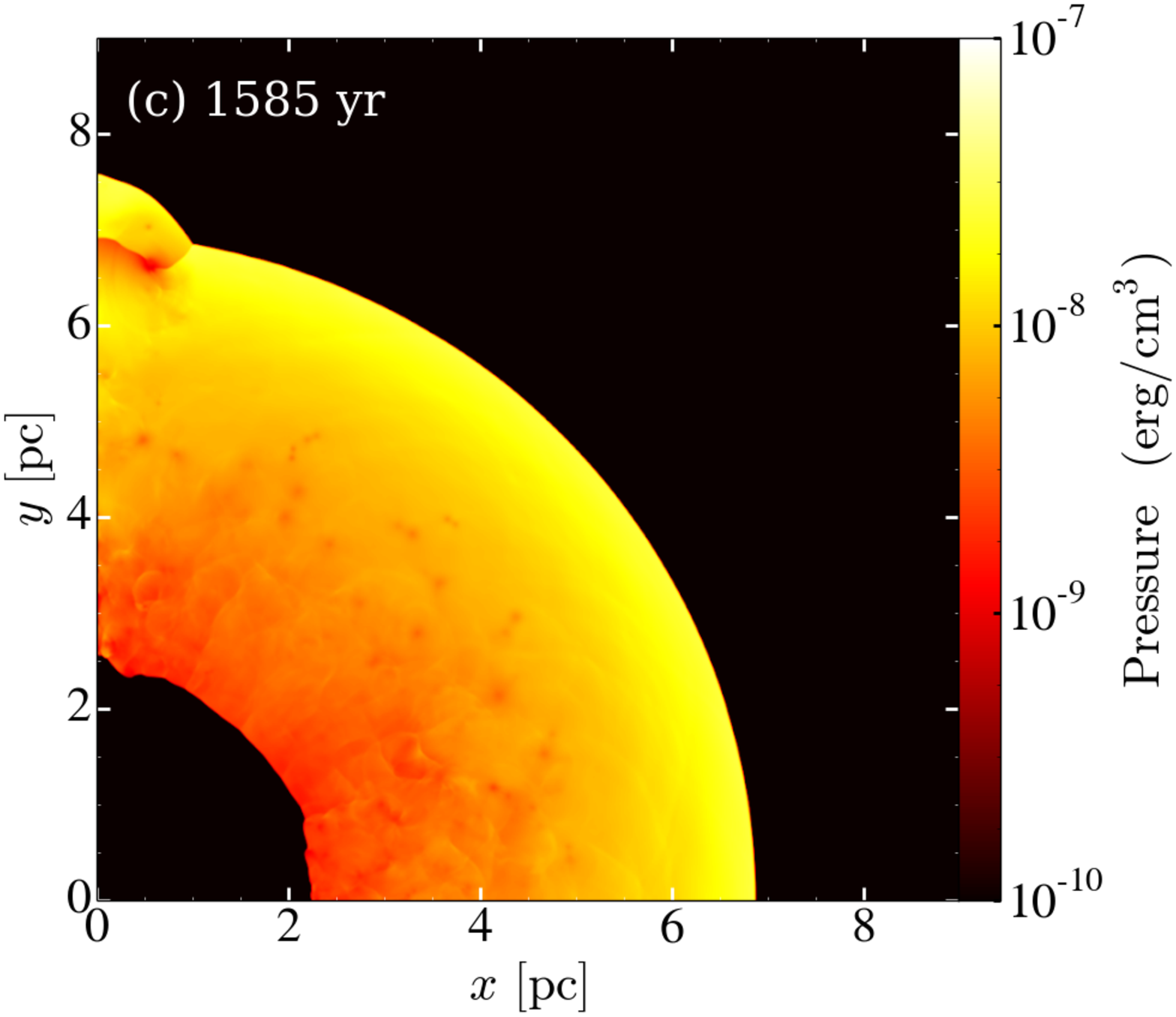}}
\subfigure{\label{subfigure:Magv2}\includegraphics*[scale=0.3,clip=true,trim=0 0 0 0]{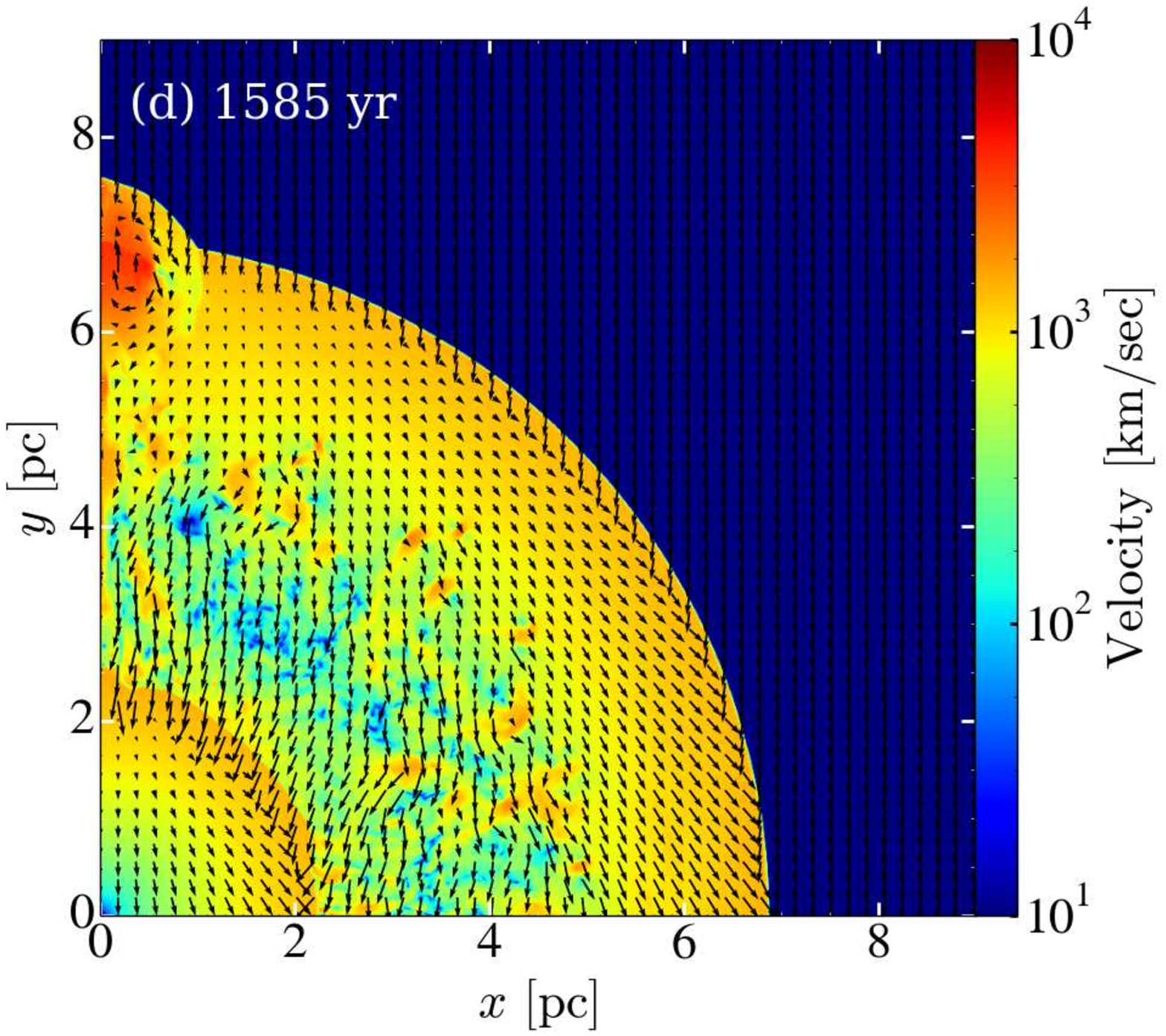}}
\caption{Same as Figure \ref{fig:results}, but the initial density of the iron clump is $\rho_{\rm blob} (r) = 1.5 \rho_{\rm ejecta}(r)$.}
\label{fig:results2}
\end{center}
\end{figure*}

\begin{figure*}
\begin{center}
\subfigure{\label{subfigure:Density3}\includegraphics*[scale=0.3,clip=true,trim=0 0 0 0]{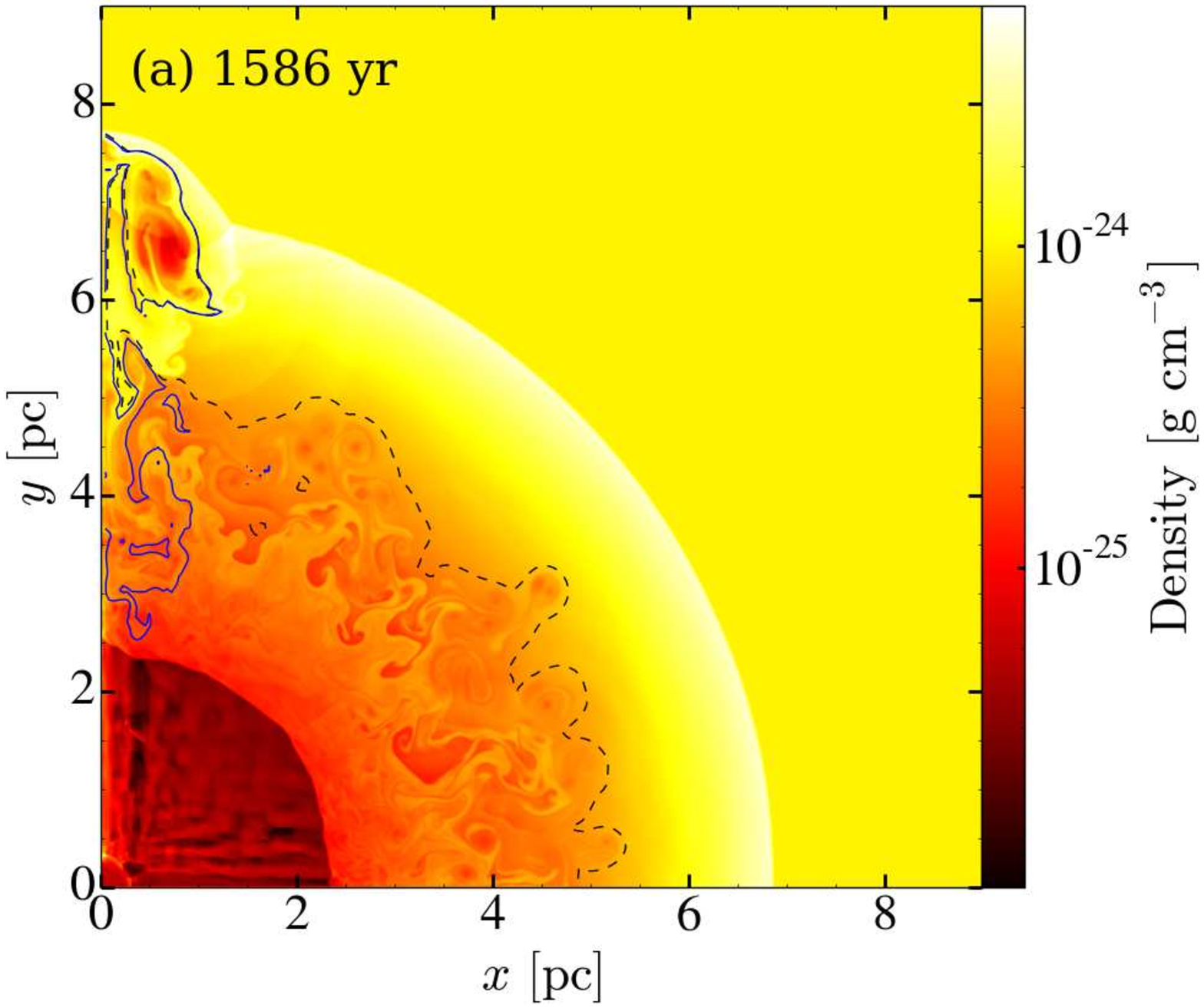}}
\subfigure{\label{subfigure:Temperature3}\includegraphics*[scale=0.3,clip=true,trim=0 0 0 0]{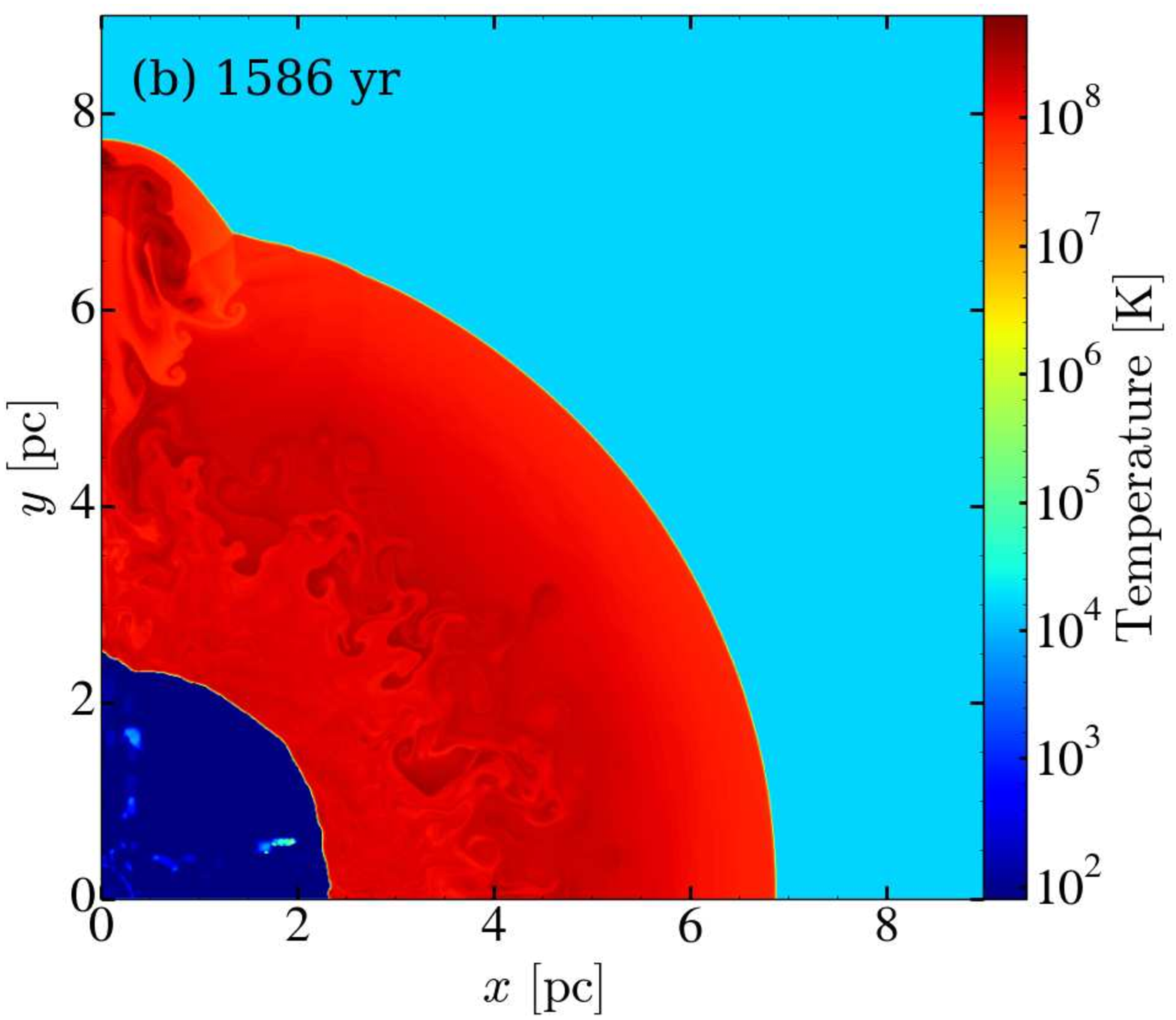}}
\subfigure{\label{subfigure:Pressure3}\includegraphics*[scale=0.3,clip=true,trim=0 0 0 0]{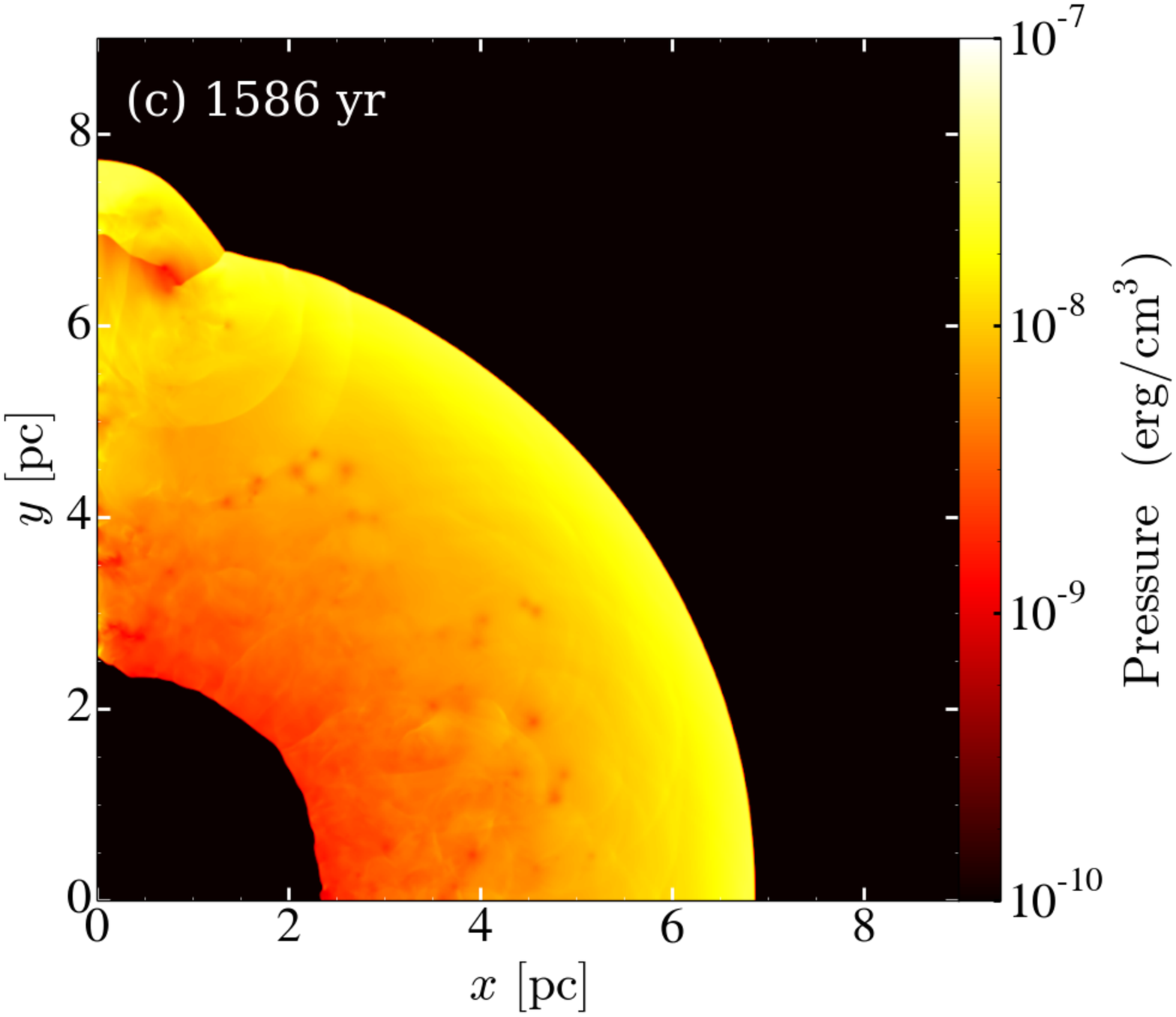}}
\subfigure{\label{subfigure:Magv3}\includegraphics*[scale=0.3,clip=true,trim=0 0 0 0]{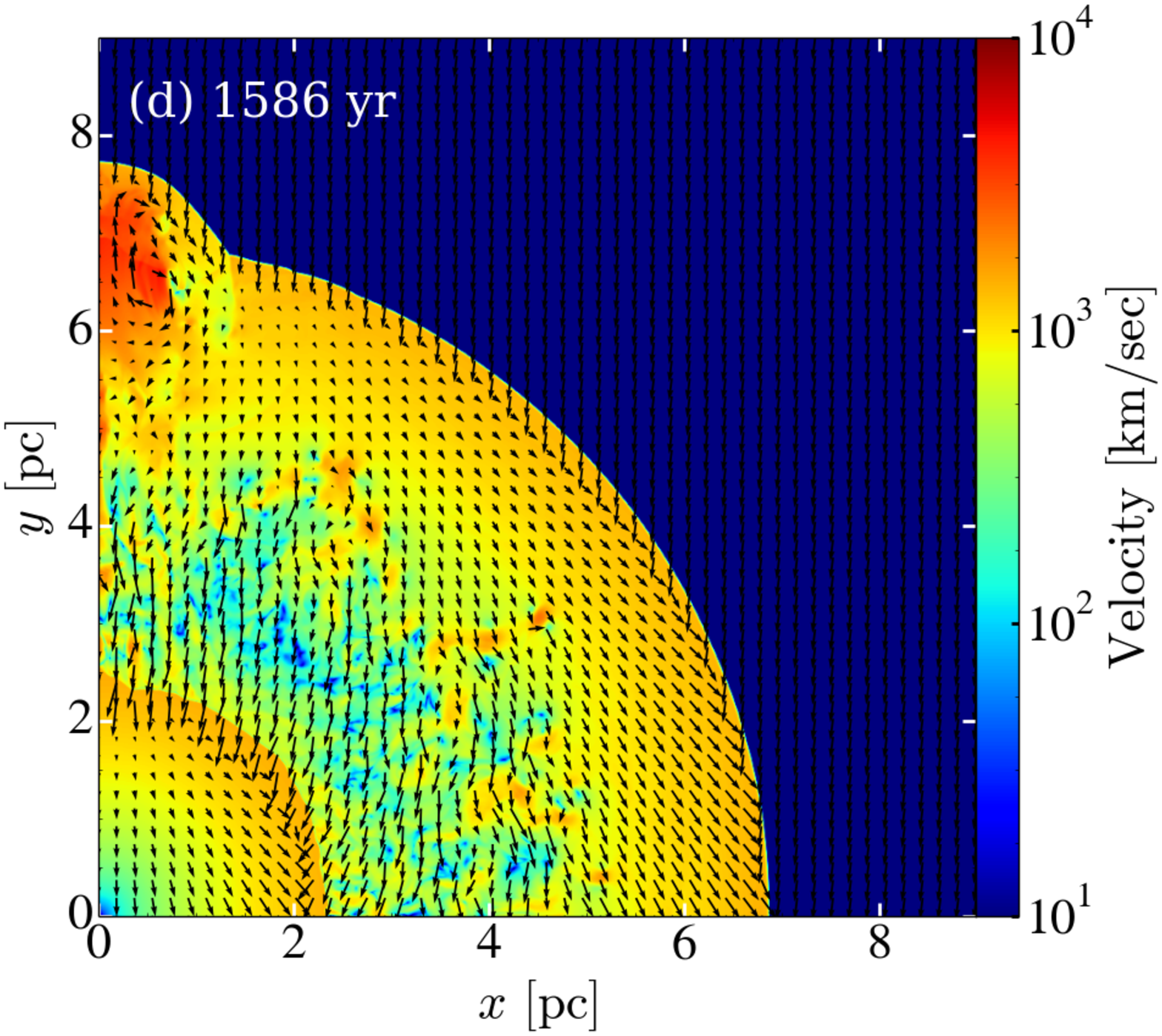}}
\caption{Same as Figure \ref{fig:results}, but the initial blob width is $D = 0.03 \pc$. }
\label{fig:results3}
\end{center}
\end{figure*}

To check the robustness of our results, we varied two parameters:
$(i)$ the iron blob density; $(ii)$ the iron blob width along the
$x$ axis. Figure \ref{fig:results2} shows the flow properties for
an iron blob having a density of 1.5 times the density of the
adjacent SN ejecta, and width of $D=0.02 \pc$ as in the fiducial
run. Figure \ref{fig:results3} shows the flow properties for an
iron blob having a density of 3 times the density of the adjacent
SN ejecta, but a wider blob with $D=0.03 \pc$.

Figures \ref{fig:results2} and \ref{fig:results3} show that
varying some of the iron blob parameters does not affect the
presence of the ear feature, only its size. Overall, our results
show that an iron blob moving alongside the SN ejecta with similar
velocity but slightly higher density (1.5-3 times the SN ejecta
density) can form a noticeable protrusion from the otherwise
spherical shape of the SNR. This might be the case in SNR~1885,
having iron tongues \citep{Fesenetal2014} and Tycho's SNR, where
metals are present inside the ears protrusions
\citep{Chiotellisetal2013}.

\section{SUMMARY}
\label{sec:summary}

Although many SNRs Ia exhibit almost spherical large-scale
symmetry, there are several SNRs Ia that show a
deviation from spherical large-scale structure. Single protrusions
exist in some SNR Ia, e.g., SNR~1885 \citep{Fesenetal2014} and Tycho
\citep{Chiotellisetal2013}, while two opposite protrusions, or 'ears', are
present in Kepler's SNR \citep{Reynolds2007} and SNR~G1.9+0.3 \citep{Reynolds2008}.
We examined a possible scenario to create such protrusions, with
2D numerical simulations of SNR Ia, as described in Section
\ref{sec:numerical}. The initial set-up includes a dense bullet,
supposed to be iron-bullet, moving with the rest of the ejecta as
described in Fig.~\ref{fig:initial}. Such iron clumps are expected
to form in the deflagration to detonation explosion model
\citep{Seitenzahletal2013}. Our results are presented in section
\ref{sec:results}. Figs.~\ref{fig:results} and \ref{fig:results0}
show the flow parameters for our fiducial case, having a blob
density 3 times the density of the adjacent SN ejecta and blob
width $D=0.02 \pc$, at two times. In Fig.~\ref{fig:results2} we
presented the results for a case with a blob having only 1.5 times
the density of the adjacent SN ejecta. In  Fig.~\ref{fig:results3}
we presented the results where the blob density is as in the
fiducial run, but twice as wide. In all cases a distinct
protrusion in the SNR shell was obtained.

In SNRs Ia having two opposite protrusions, such as
Kepler's SNR and SNR~G1.9+0.3, our scenario implies that the dense
clumps were formed along a common axis. Such a preferred axis can
result from a rotating white dwarf progenitor.
{{{ Exploring the mechanism that can produce such axis-directed 'bullets' ejecta will be at the focus of a future work,
while the present study is limited to checking the effect of such 'bullets' on the shape of the SNR.
At this point we merely take the axis-oriented clump ejecta as a work hypothesis.
}}}
If our claim holds,
this can offer an important clue to the SN Ia explosion scenario, {{{ as it requires the white dwarf progenitor to possess a preferred axis. }}} 
Namely, it {{{ might }}} require a rapidly rotating white dwarf progenitor.

The proposed process forming two opposite ears is a third
alternative to our previous two
processes \citep{TsebrenkoSoker2013}. One is the launching of jets
a long time before explosion into a circumstellar matter (CSM)
shell, namely, in a planetary nebulae phase. The SN Ia explosion
then is termed SNIP. In the second alternative the two opposite
jets are launched shortly before explosion. In the presently
studied process the ear feature is obtained without having a CSM
shell. It requires though, that the SNR interacts with a
substantial mass of ISM, such that the ejecta is slowed down. The
proposed iron bullets scenario studied here could be an
alternative process to our previously suggested ideas, lowering
the percentage of SNe Ia that originated inside PNs (SNIPs)
\citep{TsebrenkoSoker2015a}. But still, a rapidly rotating white
dwarf progenitor is required.

\section*{Acknowledgments}
This research was supported by a generous grant from the president of the Technion, Prof. Peretz Lavie.

\label{lastpage}

\end{document}